# Pelletization Characteristics of the Hydrothermal Pretreated Rice Straw with Added Binders

Xianfei Xia[1], Hongru Xiao[1]*, Zhengyu Yang[2], Xin Xie[2], Janki Bhimani[2]

1: Nanjing Research Institute for Agricultural Mechanization, Ministry of Agriculture, Nanjing, China. xiaxianfei@caas.cn, xhr2712@sina.cn

2: Dept. of Electrical & Computer Engineering, Northeastern University, Boston, USA. {yang.zhe, xie.x, bhimani.j}@husky.neu.edu

* Corresponding Author

**Abstract**: Pelletization of the loose rice straw is an attractive option to produce renewable fuels. In this paper, we focus on the problem of how to improve this pelletization process, especially to reduce energy consumption and improve product quality. In detail, we pretreat rice straw and investigate the densification characteristics of the pretreated materials. Pretreatment methods of the materials include hydrothermal treatment and adding a certain proportion of economic additives, such as rapeseed meal, waste engine oil, etc. The pretreated rice straw was pelletized by using a biomass densification platform, then the energy consumption and pellet quality were tested. Experimental results indicate that the hydrothermal pretreatment played an important role in reducing energy consumption and improving the product quality, and waste engine oil has a better effect than the rapeseed meal. We also observe that the obtained pellet quality reaches the standard of middle-grade coal, and the proposed pretreatment method realizes the comprehensive utilization of waste agricultural resources.

**Keywords**: Rice straw; Pelletization; Hydrothermal pretreatment; Energy consumption; Pellet quality; Optimization

## 1. Introduction

With the growing global energy and environment problems, clean and renewable biomass fuels are paid more attention by many countries. China is an agricultural country which has over 20 types of crop straws, and the harvest yield is more than 7 tons each year. However, most of these abundant agricultural biomass resources are abandoned or uncontrolled burned. To eliminate the current energy and environment problems, it is required to make a good use of these biomass resources. Among the various energy utilization ways of crop straw, pelletization is a process which has high productivity [1-3], high density and high strength. The biomass fuels produced in this way is also convenient for transportation and burning[4, 5]. As a result, the biomass pelletization has gradually become the mainstream technology. However, the application of the rice straw pelletization technology has two technical bottlenecks, high energy consumption and low product calorific value.

To improve the biomass pelletization process, Kong *et al.* [6] discovered that the product quality could be improved by adding linen fibers. Rosin *et al.* [7] mixed biomass materials with various additives such as starch, molasses, HSC-residue, slaked lime, ash and montan resin, to improve the product density. Mediavilla *et al.* [8] found that the physical quality of the biomass pellet was higher when lignosulphonate was added. To increase the calorific value of rye straw pellets, paraffin, palmitin, anthracite coal and lignite coal were mixed in the biomass material [9]. Shang *et al.* [10] investigated the influence of rapeseed oil on the friction during densification process. Zarringhalam *et al.* [11] added coal waste into the biomass material and found that the water resistance index and tensile strength of the pellets were improved. Zannikos *et al.* [11][12] mixed sawdust with waste plastic to produce pellets, and analyzed the calorific value and emissions of the products. Filbakk *et al.* [13] studied the effect of bark content on the quality of the scots pine pellets. Liu *et al.* [14] mixed rice straw with bamboo to make pellets.

At present, torrefaction and hydrothermal carbonization are the two main pretreatment methods [15-20]. Acharya *et al.* [21] and Pirraglia *et al.* [22] made a comprehensive economy evaluation of the biomass torrefaction technology, and pointed out that torrefaction was an important approach to promote the efficient utilization of biomass resources. Stelte *et al.* [23, 24] discovered that torrefaction could increase the energy consumption and reduce the quality of the products. Liu *et al.* [25] carried out experiments and found that hydrothermal treated biomass pellets had higher carbon content and calorific value, and lower ash content, but the compressive strength of the product decreased. Reza *et al.* [26, 27] used hydrothermal pretreated biomass to produce pellet, and found that both calorific value and physical quality of the product were improved.

The pretreated biomass material is more homogeneous, which makes it is easier to be pelletized and the volumetric energy density could be improved. Therefore, the raw material pretreatment technology would be a worthy research area for rice straw utilization. However, recent researches mainly focus on wood-biomass, and the pretreatment way was usually single, just physical or simple chemical pretreatment. There are few researches related to rice straw material or composite pretreatment. Meanwhile, the material pretreatment had composite influences on energy consumption, die wear and product quality. Therefore, the rice straw was pretreated by a composite pretreatment way which included hydrothermal carbonization treatment and adding economic additives such as waste engine oil and rapeseed meal firstly in this paper. Then the pretreated material was pelletized, and energy consumption and pellet quality were tested. Last, the optimization scheme was selected. The purpose of this research is to reduce energy consumption of the rice straw densification process, and increase the calorific value of the products by making best posible use of the waste resources.

## 2. Experiment and methods

### 2.1 Materials and experimental design

The raw material was rice straw harvested in Jiangyan, Jiangsu province, China in the autumn of 2016. Rice straw was dried naturally and then grinded to 2 mm, the granularity distribution is shown in Figure 1. Hemicellulose content of the experimental rice straw is 27.2%, cellulose content is 34%, lignin content is 14.2%, the stacking density is 75.3 kg/m$^3$, and the calorific value is 14.753 MJ/kg.

Experiments in this paper mainly focus on the influence of different pretreatment method on energy consumption and product quality of the rice straw pelletization process. Firstly, the non-pretreated rice straw was pelletized. Then, the selected binders were added into the raw materials and the mixtures were pelletized. Finally, the rice straw was hydrothermal pretreated, and the selected binders were mixed into the pretreated rice straw and the mixtures were pelletized. The raw materials were pelletized by a specific pelleter platform which has the function of pelletization, testing energy consumption and tensile strength.

### 2.2 Selection of additives

The additives should be cheap and will improve the physical and chemical properties of the raw material, increase the calorific value of the product, and reduce the friction during the densification process to reduce energy consumption. Furthermore, the additives should neither have bad influence on the quality of the products nor cause environmental pollution. Particularly, the waste needed to be dealt with is preferred. Therefore, waste engine oil and rapeseed meal are chosen as the additives initially, the main combustion products of these additives are $CO_2$ and $H_2O$. The additive amount is arranged between 1% and 8%. Then more, starch was chosen as adhesive and its amount is arranged as 4%.

**Waste engine oil**

Lubricating oil is used for lubrication, rust prevention, sealing and cooling in various types of vehicles and mechanical equipment, and the engine oil is a kind of lubricating oil used in engines. Its main components are hydrocarbons, and the calorific value is more than 30 MJ/kg. As the engine oil must be replaced when it is used after a certain period, a large amount of waste engine oil will be produced every year. Making good use of this waste engine oil can not only make waste profitable and save energy, but also protect the environment. The waste engine oil sample is shown in Figure 2.

**Rapeseed meal**

Rapeseed is one of the most important oil crops in China, and the annual yield of rapeseed is more than 11 million tons. Rapeseed is mainly used to make rapeseed oil, and 350~400 kg oil can be produced from 1 ton rapeseed, then plenty of rapeseed meal is left over. Although rapeseed meals are rich in protein, but they are harmful because of anti nutritional ingredients contained in rapeseed meal such as thioglycoside, erucic acid, phytic acid and polyphenol. This limits its use as protein source for animal feed. Therefore, only a few rapeseed meals could be added into the feed and most of them are used as organic fertilizer directly. In fact, the content of oil and lignin is very high in rapeseed meal, and its calorific value is about 19.9 MJ/kg [28]. In this experiment, rapeseed meal is chosen as one kind of additives. The rapeseed meal sample is shown in Figure 2.

**Starch**

Starch is a renewable carbohydrate which contains a small amount of fat and protein and is widely used as adhesives in biomass densification process [29]. It has been discovered that the quality of the pellets is improved when starch is added into the raw material.

**2.3 Hydrothermal pretreatment**

Hydrothermal pretreatment is also called auto-hydrolysis. The water is kept subcritically or critically under a high pressure by increasing the temperature. And the stable lignocellulose in biomass materials would be hydrolyzed efficiently in the hydrothermal process. Hydrothermal pretreatment takes water rather than any other chemical reagents as medium. This process is controllable and has short reaction period, which is friendly to the environment. Based on the difference of processing time and temperature, hydrothermal pretreatment is divided into hydrothermal gasification, hydrothermal liquefaction and hydrothermal carbonization (HTC). Compared with hydrothermal gasification and hydrothermal liquefaction, hydrothermal carbonization needs lower temperature and pressure, and the reaction condition is moderate. In the hydrothermal carbonization process, the temperature is kept at 180~250 °C and the pressure is about 1.4~27.6 MPa. The products of hydrothermal carbonization are uniform-sized and have high energy density [30].

Biomass hydrothermal carbonization would produce solid, liquid and gaseous products. The solid product is coke, the liquid products include reducing sugar, acetic acid and TOC, and the gaseous product is mainly $CO_2$. In this experiment, the rice straw is treated by hydrothermal carbonization, and the solid products are pelletized. Temperature of the pretreatment is 240 °C and the reaction time is 30 min.

**2.4 Testing apparatus and process**

The pelleter device used in this paper is a self-made pelletization platform. This platform is made up of two parts: a pelletization apparatus and an electronic universal testing machine. The pelletization apparatus can simulate the real pelletization environment. A heating element system which can heat the die quickly was adopted in the apparatus. The electronic universal testing machine is produced in HRJ Company in Jinan, Shandong province, China. The maximum test load is 100 kN and the extrusion speed is 0.01~1000 mm/min. The pelletization platform is shown in

Figure 3.

The testing process include the following, (1) adjusting the pelleter apparatus to make sure the pressing shaft is aligned with the cavity of the die; (2) heating the die to the target temperature, and then adding the weighed material into the die; (2) compressing the materials at a certain speed until the predetermined pressure is reached; and (4) maintaining the pressure for some time and finally extruding the pellet out of the die.

The control computer was used to record the stress, displacement and deformation data during the whole process. In this experiment, the length to diameter ratio of the die is 4, the temperature is 120 °C, the compressing speed is 50 mm/min, the pressure is 60 MPa, and the pressure holding time is 60 s. Moisture content of the pelletization materials is 10%.

**2.5 Experiment testing indicators**

The pelletization platform shown in Figure 3, was used to compress the rice straw to make pellets. And then several tests were carried out to evaluate the densification process. The testing items include specific energy consumption, tensile strength and pellet density; moreover, calorific value was also obtained through experiment.

2.5.1 Specific energy consumption

Specific energy consumption means the energy consumption of per unit mass biomass when it compressed from loose form to densified pellet, it is an important evaluation index of the pelletization process. The displacement-stress curve recorded by the computer can be used to calculate the energy consumption, and then the specific energy consumption can be calculated by formula (1).

$$E = \frac{W}{m} = \frac{S}{m}\int_0^l \sigma(x)dx \tag{1}$$

*Where,* $E$ is specific energy consumption, J/kg; $W$ is the total energy consumption of a single pellet, J; $m$ is the weight of the pellet, g; $\sigma$ is the pelleting pressure, Pa; $x$ is the displacement, m; $l$ is the maximum compressive displacement, m; $S$ is the cross-section area of the pressing shaft, m$^2$.

2.5.2 Tensile strength

Tensile strength is used to evaluate the crushing resistance and mechanical stability of the pellets. The pellet with a certain length was placed on the pelletization platform transversely, and the pressing shaft pressed it at a certain speed until the pellet was broken, the maximum compressing force $F$ was record. The tensile strength is calculated by formula (2).

$$\sigma_t = \frac{2F}{\pi dl} \tag{2}$$

*Where,* $\sigma_t$ is the tensile strength, Pa; $F$ is the compressing maximum force, N; $d$ is the diameter of the pellet, m; $l$ is the length of the pellet, m.

2.5.3 Relaxed density

Relaxed density of the pellet is measured based on the standard of NY/T 1881.1-2010 (Densified biofuel – Test methods Part 1: General principle, China). The sample was cooled for 24 h in the air, and then the mass of the pellet was measured by an electronic scale, the length and diameter of the pellet were also measured by a vernier caliper. The density of the sample is calculated by formula (3).

$$\rho = m/V_p = \frac{4m}{\pi DL} \tag{3}$$

Where, $\rho$ is the relaxed density, g/cm$^3$; $m$ is the mass of the pellet, g; $D$ is the diameter of the pellet, cm; $L$ is the length of the pellet, cm.

### 2.5.4 Calorific value

The apparatus used to test the calorific value is an automatic rapid calorimeter, which is made by Shanghai OURUI instruments Equipment co., LTD. A weighed pellet was taken as the sample, and then a certain length of conductive wire was convolved around the pellet. Connecting the sample with the oxygen bomb support through the conductive wire, finally, installing the oxygen bomb filled with oxygen into the calorimeter and measuring the calorific value.

## 3. Results and analysis

The non-pretreated and pretreated rice straw pellet is shown in Figure 4. It can be seen from Figure 4 that color of the pellet becomes darker after hydrothermal carbonization (HTC) treated. This is mainly due to the increasing content of carbon after hydrothermal carbonization treated. Figure 5 is the compaction curves of the non-pretreated and pretreated rice straw; the specific energy consumption is calculated according the curves. It should be noted that the hydrothermal pretreated straw is difficult to be compressed into pellets without binders.

According to the testing methods introduced in section 2, several parameters including specific energy consumption (SEC), relaxed density (RD), tensile strength (TS) and calorific value (CV) of the products were obtained. The testing results are shown in Table 1.

### 3.1 Specific energy consumption

In Table 1, it can be found that the specific energy consumption of the untreated rice straw is 50.089 J/g, the energy consumption of the pretreated rice straw is between 19.492 J/g and 47.392 J/g, it is obviously that the energy consumption decreased when the materials was pretreated. Among all the pretreated methods, the HTC pretreated rice straw with the additives of 10% waste engine oil and 4% starch performed the best; its specific energy consumption was only 19.492 J/g. The influence of each pretreated method on the specific energy consumption was shown in Figure 6 based on the data in Table 1.

Figure 6 showed the influence of different pretreated method on the specific energy consumption of the non-pretreated and pretreated straw. The specific energy consumption gradually decreased with the increase of additive proportion of the additives. In the two types of additives, waste engine oil performed better than rapeseed meal in saving energy. And the specific energy consumption of hydrothermal pretreated straw was lower than that of the untreated straw while the additives were the same.

**3.2 Relaxed density of the pellet**

Relaxed density of the pellet was measured based on the method in section 2, and the testing results were shown in Table 1. It can be seen from Table 1 that the relaxed density of the untreated straw was 1.051 g/cm$^3$, the relaxed density of the composite pretreated straw was between 0.975 g/cm$^3$ and 1.268 g/cm$^3$. According to the agricultural trade standard of NY/T 1878-2010 (Specification for Densified Biofuel) in China, density of the herbaceous granular biomass solid fuel should not less than 1000 kg/m$^3$. In other words, the product was qualified only if the density is over 1 g/cm$^3$. Based on the obtained data in Table 1, the influence of different pretreated ways on density was shown in Figure 7. The dotted line in Figure 7 was the density qualification line. If the density is above the dotted line, the products are qualified and the corresponding pretreatment method is feasible. Otherwise, the products are not qualified and the corresponding pretreatment method is infeasible.

Figure 7 showed the influence of different additives on the density of pellet made from untreated materials. When the additive was rapeseed meal, the density increased with the increase of additive proportion. For the waste engine oil, the density decreased with the increase of additive proportion. Rapeseed meal performed better in improving the density. It is observed that the relaxed density, of the pellet made by hydrothermal pretreated straw is improved significantly compared to untreated straw. The pellet density failed to meet the standard if the additive proportion of waste engine oil was more than 9%. Thus, some of the pretreatment ways could improve the pellet density, and some would decrease it.

**3.3 Tensile strength of the pellet**

Tensile strength of the pellet was measured based on the method in section 2, and the testing results were shown in Table 1. It can be seen from Table 1 that the tensile strength of untreated straw pellet was 1.130 MPa, tensile strength of pretreated straw was between 0.612~2.806 MPa. Lu *et al.* [31] pointed out that tensile strength of the straw pellet biofuel shouldn't be lower than 0.81 MPa. Thus, 0.81 MPa was taken as the standard of the tensile strength of rice straw pellet in this paper. Based on the obtained data in Table 1, the influence of different pretreated ways on the tensile strength of the pellet was shown in Figure 8. In Figure 8, If the tensile strength is above the dotted line, the pellet is qualified and the corresponding pretreated method is feasible, otherwise, the product is not qualified and the corresponding pretreatment method is infeasible.

Figure 8 showed the influence of different pretreated method on the tensile strength of the pellet.

When the additive was rapeseed meal, the tensile strength increased with the increase of additive proportion significantly. The tensile strength of the pellet made from hydrothermal pretreated straw decreased compared to untreated straw. For the waste engine oil, the tensile strength decreased with the increase of additive proportion. Rapeseed meal performed better in improving the tensile strength, when the waste engine oil was more than 8%, tensile strength of the pellet failed to meet the standard. Thus, some of the pretreatment ways could improve the tensile strength, and some would decrease it.

**3.4 Calorific value**

Calorific value of the pellet was measured based on the method in section 2, and the testing results were shown in Table 1. It can be seen from Table 1 that the lower calorific value of the untreated straw pellet is 14.753 MJ/kg and the lower calorific value of the pretreated straw pellet was between 15.026 MJ/kg and 25.113 MJ/kg. We observe that the lower calorific value increased with the increase of the additive proportion of additives. Waste engine oil performed the better in improving the lower calorific value. It was concluded that the lower calorific value could improve a lot by hydrothermal pretreatment. For example, when additive proportion of waste engine oil was 4%, the lower calorific was improved 11.96%, but for the hydrothermal pretreated straw, the lower calorific was improved 67.29% compared to untreated straw pellet. Therefore, all the pretreated method used in this paper could improve the lower calorific value of the pellet effectively.

**3.5 Optimal selection and analysis**

3.5.1 Selection of the optimal pretreated method

The qualified rice straw pellet should have proper density, high tensile strength and calorific value, meanwhile, the energy consumption of the densification process should be low. Based on the analysis of the testing results above, four preferable pretreated methods were selected out. The selected pretreated methods included untreated rice straw with 8% waste engine oil (G35) addition and hydrothermal pretreated rice straw with 7% waste engine oil and 4% starch (SR07D4) addition. The selected methods had better comprehensive pellet quality and lower energy consumption. Comparison of testing items between the selected pretreated methods and the untreated rice straw (D1) was listed in Table 2.

We observe from Table 2 that the energy consumption of the selected pretreated scheme G35 was 47.76% less than that of the untreated rice straw, and pellet density was decreased 1.71%, the tensile strength was improved 2.83%, meanwhile the calorific value was improved 15.24%. However, the energy consumption of selected pretreated scheme SR07D4 was 51.08% less than that of the untreated straw, the pellet density was improved 2.66%, the tensile strength was decreased 23.01%, and the calorific value was improved 68.90%. Therefore, the comprehensive improvement effect of the scheme SR07D4 was the best among the four selected methods.

## 4 Conclusions

In this paper, we pretreated the rice straw by the hydrothermal carbonization method which uses the rapeseed meal and waste engine oil as additives. The densification energy consumption and pellet quality of the pretreated materials were tested. The following conclusions were drawn:

1) Densification energy consumption of the rice straw was gradually decreased with the increase of additive proportion of additives. Waste engine oil performed better than rapeseed meal in saving energy. Among all the pretreated methods, the HTC pretreated rice straw with the additives of 10% waste engine oil and 4% starch performed the best.

2) Relaxed density of the rice straw pellet increased with the increase of additive proportion of rapeseed meal, and the density decreased with the increase of additive proportion of waste engine oil. And relaxed density of the pellet made by hydrothermal pretreated straw was improved significantly compared to the untreated straw.

3) Tensile strength of the rice straw pellet increased with the increase of additive proportion of rapeseed meal, and the tensile strength decreased with the increase of additive proportion of waste engine oil. Tensile strength of the pellet made from hydrothermal pretreated straw decreased compared to untreated straw. Some of the pretreatment ways could improve the tensile strength, and some would decrease it.

4) Calorific value of the pretreated rice straw pellet increased with the increase of the additive proportion of additives effectively. Calorific value of the pellet improved significantly by hydrothermal pretreatment.

5) The rice straw pretreated by hydrothermal carbonization with the addition of 7% waste engine oil and 4% starch has good performance. The energy consumption was 51.08% less than that of the untreated straw, the pellet density was improved 2.66%, and the calorific value was improved 68.90%, while the tensile strength was decreased 23.01% only. Combustion characteristic of this kind of pellet was excellent, it would be used as a good substitute fuel for coal.

## Acknowledgments

This work was supported by The Natural Science Foundation of Jiangsu Province (BK2011706), and The Prospective Industry-Study-Research Cooperation Foundation of Jiangsu Province (BY2012023).

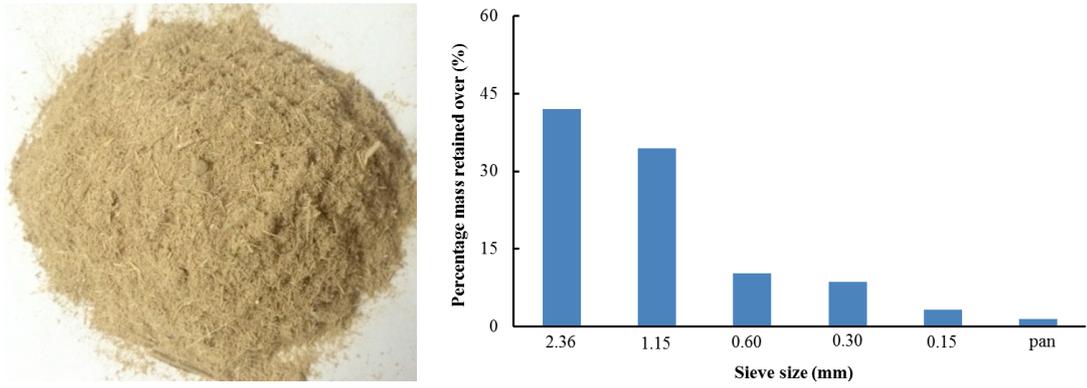

Fig.1 Granularity distribution of the grinded rice straw

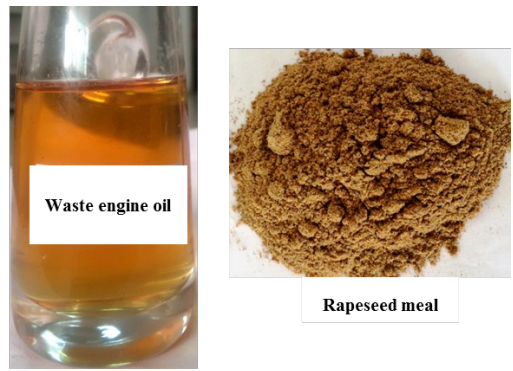

Fig.2 The additives samples (waste engine oil and rapeseed meal)

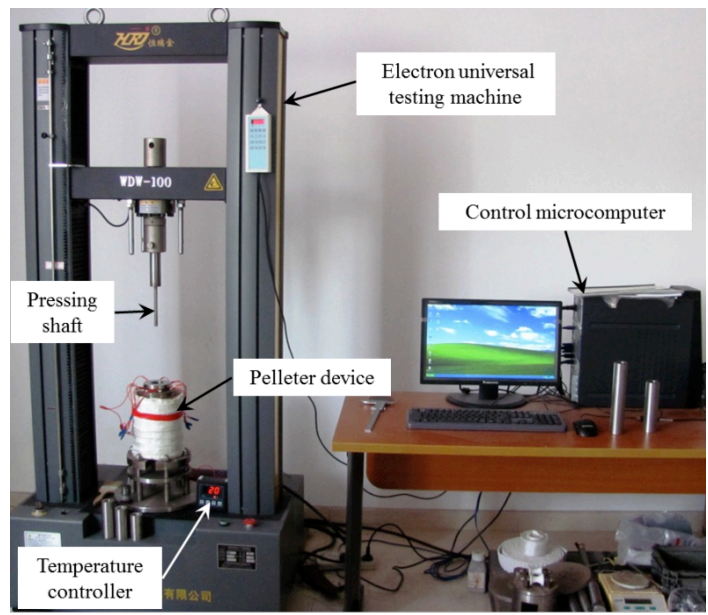

Fig.3 The pelletization platform

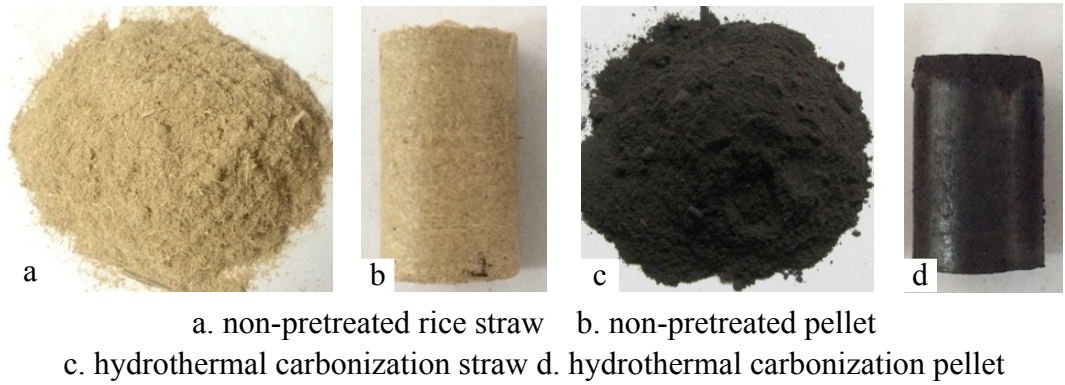

a. non-pretreated rice straw     b. non-pretreated pellet
c. hydrothermal carbonization straw d. hydrothermal carbonization pellet
Fig.4 Non-pretreated, hydrothermal carbonization rice straw and the corresponding pellet

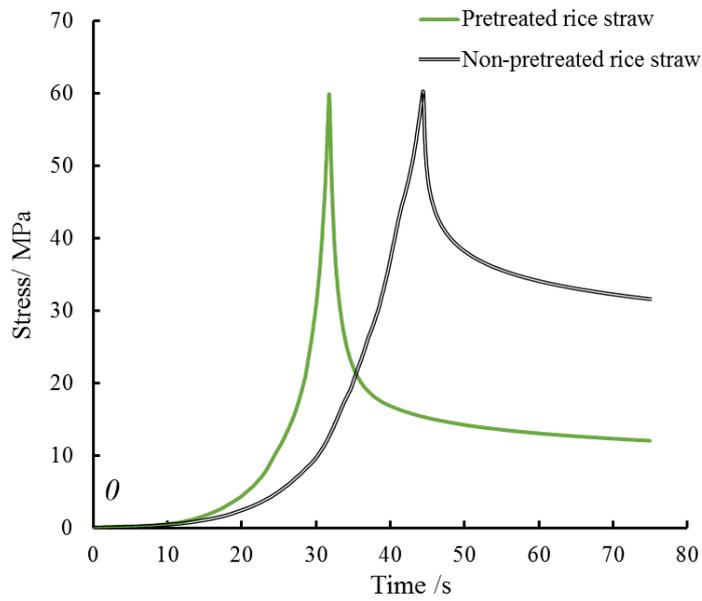

Fig.5 Compaction curves of the non-pretreated and pretreated rice straw

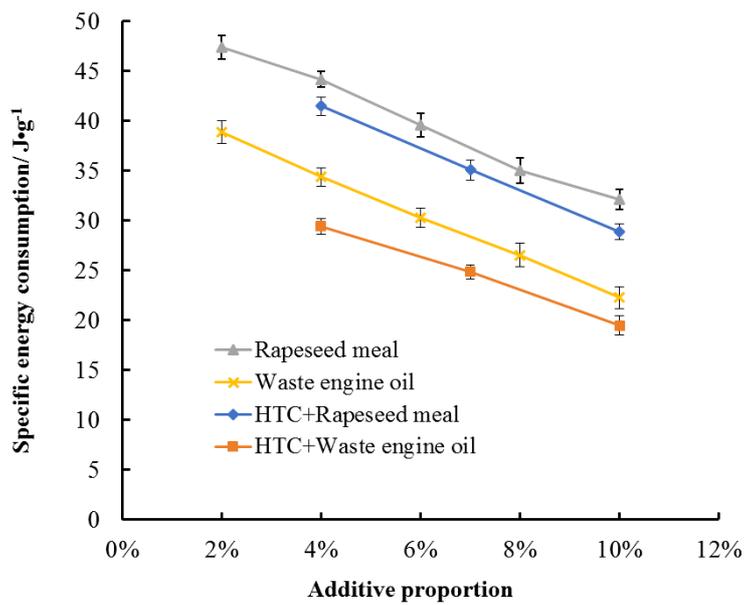

Fig.6 The influence of different pretreated method on the specific energy consumption (SEC)

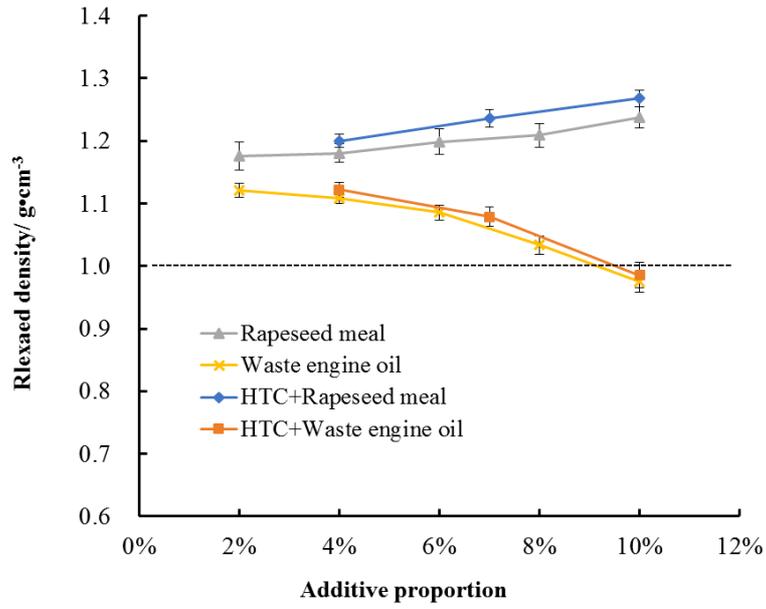

Figure 7 The influence of different pretreated method on the relaxed density (RD)

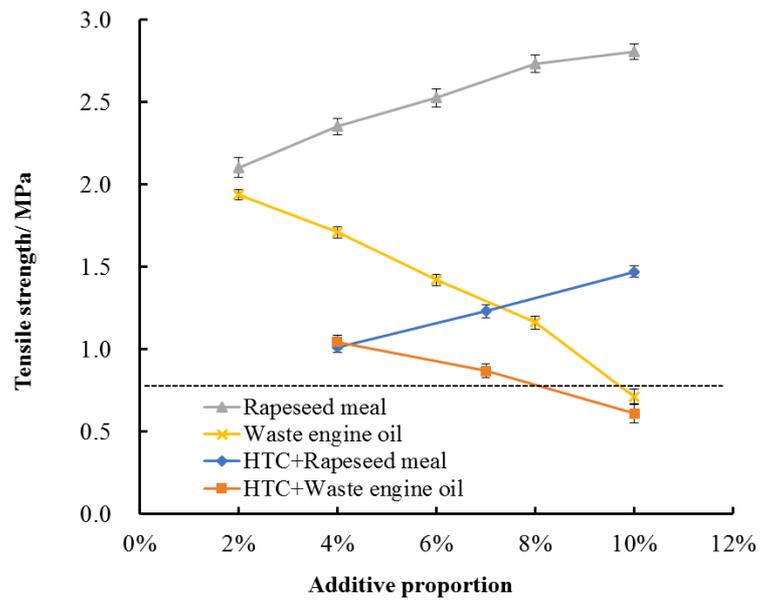

Fig.8 The influence of different pretreated method on the tensile strength (TS)

Tab. 1 Testing results for the rice straw pellet pretreated in different ways

| Test NO. | Serial NO. | Pretreated method | Additives | Content | SEC (J/g) | RD (g/cm$^3$) | TS (MPa) | CV (MJ/kg) |
|---|---|---|---|---|---|---|---|---|
| 1 | D1 | None | None | None | 50.089 | 1.051 | 1.130 | 14.753 |
| 2 | G32 | | | 2% | 47.392 | 1.176 | 2.103 | 15.026 |
| 3 | G33 | | | 4% | 44.172 | 1.180 | 2.351 | 15.147 |
| 4 | G34 | None | Rapeseed meal | 6% | 39.565 | 1.199 | 2.526 | 15.257 |
| 5 | G35 | | | 8% | 35.017 | 1.209 | 2.734 | 15.421 |
| 6 | G36 | | | 10% | 32.138 | 1.238 | 2.806 | 15.650 |
| 7 | G42 | | | 2% | 38.874 | 1.121 | 1.939 | 16.106 |
| 8 | G43 | | | 4% | 34.375 | 1.108 | 1.710 | 16.518 |
| 9 | G44 | None | Waste engine oil | 6% | 30.284 | 1.086 | 1.421 | 16.889 |
| 10 | G45 | | | 8% | 26.541 | 1.033 | 1.162 | 17.002 |
| 11 | G46 | | | 10% | 22.293 | 0.975 | 0.712 | 17.517 |
| 12 | SR04D4 | | | 4%+4%D | 29.427 | 1.122 | 1.045 | 24.681 |
| 13 | SR07D4 | Hydrothermal pretreatment | Waste engine oil | 7%+4%D | 24.857 | 1.079 | 0.870 | 24.918 |
| 14 | SR10D4 | | | 10%+4%D | 19.492 | 0.985 | 0.612 | 25.113 |
| 15 | SC04D4 | | | 4%+4%D | 41.472 | 1.200 | 1.012 | 20.141 |
| 16 | SC07D4 | Hydrothermal pretreatment | Rapeseed meal | 7%+4%D | 35.074 | 1.235 | 1.231 | 20.646 |
| 17 | SC10D4 | | | 10%+4%D | 28.879 | 1.268 | 1.472 | 21.173 |

Note: In Table 1, "D" means the content of starch. For example, "4%D" means the adding mass fraction of starch is 4%.

Table 2 Comparison of testing items between the selected pretreated methods

| Item NO. | Serial NO. | SEC (J/g) | Variation range | RD (g/cm$^3$) | Variation range | TS (MPa) | Variation range | CV (MJ/kg) | Variation range |
|---|---|---|---|---|---|---|---|---|---|
| 1 | D1 | 50.089 | — | 1.051 | — | 1.130 | — | 14.753 | — |
| 2 | G35 | 26.541 | -47.76% | 1.033 | -1.71% | 1.162 | +2.83% | 17.002 | +15.24% |
| 3 | SR07D4 | 24.857 | -51.08% | 1.079 | +2.66% | 0.870 | -23.01% | 24.918 | +68.90% |